\documentclass[aps,floatfix,superscriptaddress,notitlepage,nofootinbib]{revtex4-1}
\usepackage{amsmath,amssymb,amsfonts,graphics,graphicx,dcolumn,bm,enumerate}
\usepackage{comment,natbib,appendix}
\usepackage{multirow,color}
\usepackage{chngpage}
\usepackage{afterpage}
\usepackage{xcolor}
\usepackage{amsthm}
\usepackage{natbib}
\usepackage{hyperref}
\usepackage[margin=1.2in]{geometry}
\usepackage{epstopdf}
\usepackage{float}
\usepackage{amsmath}

\newcommand{\cmp}
{\affiliation{Saha Institute of Nuclear Physics, Kolkata 700064, India.}}
\newcommand{\isi}
{\affiliation{Economic Research Unit, Indian Statistical Institute, Kolkata 700108, India.}}
\newcommand{\raghunathpur}
{\affiliation{Department of Physics, Raghunathpur College, Raghunathpur, Purulia 723133, India.}}

\begin{document}
\title{Do  Successful Researchers  Reach the Self-Organized Critical Point?}

\author{Asim Ghosh}
\email[Email: ]{asimghosh066@gmail.com}
\raghunathpur
 
 \author{Bikas K. Chakrabarti }%
 \email[Email: ]{bikask.chakrabarti@saha.ac.in}
 \cmp \isi 
 
\begin{abstract}
The index of success of the researchers is now mostly measured
using the Hirsch index ($h$). Our recent precise demonstration,  that statistically $h \sim \sqrt {N_c}
\sim \sqrt {N_p}$, where $N_p$ and $N_c$ denote respectively  the total
number of publications and total citations for the researcher, suggests
that average number of citations per paper ($N_c/N_p$), and hence $h$, are
statistical numbers (Dunbar numbers) depending on  the community or
network to which the researcher belongs. We show here, extending our
earlier
observations, that the indications of success are not reflected by the
total citations $N_c$, rather by the inequalities among citations from
publications to publications. Specifically, we show that for very successful
authors, the yearly variations  in the Gini index ($g$, giving the average
inequality of citations for the publications) and the Kolkata index ($k$,
giving the fraction of total citations received by the top $1 - k$
fraction of publications; $k = 0.80$ corresponds to Pareto's 80/20 law)
approach each
other to $g = k \simeq 0.82$,  signaling a precursor for the arrival of (or
departure from) the Self-Organized Critical (SOC) state of his/her
publication statistics. Analyzing the citation statistics (from Google
Scholar) of thirty successful scientists throughout their recorded
publication history, we  find that the  $g$ and $k$ for very successful
among them (mostly Nobel Laureates, highest rank Stanford Cite-Scorers,
and a few others) reach and hover just above (and then) below that $g = k
\simeq 0.82$ mark,  while for others they remain below that mark. We also
find that  all the lower (than the SOC mark 0.82) values of $k$ and $g$
fit a linear relationship $k = 1/2 + cg$, with $c = 0.39$, as suggested by
an
approximate Landau-type expansion of the Lorenz function, and this also
indicates $k = g \simeq 0.82$ for the (extrapolated) SOC precursor mark.
\end{abstract}

\maketitle


\section{Introduction}
Inspiring researches in sociophysics (see e.g. \cite{1Oliveira1999,2Chakrabarti2006,3Castelano2009,4Helbing2010,5Galam2012,6Sen2014}) have, in years, led
to intense research activities in several statistical and  statistical
physical models and analysis of socio-dynamical problems.  For example,
the social opinion formation models of Galam (see e.g., \cite{7Galam2008,8Galam2020}), of
Biswas-Chatterjee-Sen (see e.g. \cite{9Biswas2023,10Filho2023}), of Minority Games (see e.g.
\cite{11Challet2005}), of Kolkata Paise  Restaurant games (see e.g., 
 \cite{12Martin2019,13Harlalka2023}), etc. 
In view of the automatically encoded
wide range of the citation data of the
publications by the scientists  and their
easy availability in the internet, we
have studied here the inequality statistics
from Google Scholar data.  The presence of
ubiquitous inequalities allowed recently the
studies of various scaling etc properties in
their statistics (see e.g., \cite{14Ghosh2022,15Ghosh2023}) of the
Hirsch index \cite{16Hirsch2005}, or the universal (or
limiting) Self-Organized Critical (SOC)
behavior (see e.g., \cite{17Manna2022,18Ghosh2021,19Banerjee2023}) and their
citation inequality like the century-old
Gini ($g$) \cite{20Gini1921} and the recently
introduced Kolkata  ($k$) \cite{21Ghosh2014,22Banerjee2020} indices.
It may be noted at this stage that while Gini ($g$) values
measure the overall inequality in the distributions and the
Kolkata index ($k$) gives the fraction of ``mass" or of total citations
coming from the $(1 - k)$ fraction of avalanches or  publications.
These studies \cite{17Manna2022,18Ghosh2021,19Banerjee2023} indicated that the inequalities in the avalanche
size distributions, measured by $g$ and $k$, just prior to the arrival of
the SOC point in several standard physical models (like the sand-pile
models of Bak–Tang–Wiesenfeld \cite{23Bak1987}, Manna \cite{24Manna1991}, and others), and in social
contexts of  citations from publications \cite{18Ghosh2021,19Banerjee2023} becomes equal ($g = k
= 0.84 \pm 0.04$). It may also be noted that $k = 0.80$ corresponds to
Pareto's 80/20 law (see e.g., \cite{21Ghosh2014,22Banerjee2020}). This Pareto Principle asserts
that 20\% of the causes are responsible for 80\% of the outcomes. In other
words, the principle suggests that a small fraction of the factors contribute
in causing a large fraction of major events, from economics to quality
management and even in personal development. In business, it is often
used to identify the most important areas for improvement. It may be mentioned here that our earlier studies of inequality indices $g$ and $k$ \cite{17Manna2022,18Ghosh2021,19Banerjee2023,20Gini1921,21Ghosh2014,22Banerjee2020} corresponded to the
cumulative dynamics (as the sand-pile dynamics progresses and cluster
distributions grow or the publications by the authors or from the
institutions progresses over time and the citation size distributions
grow since the start of the dynamics) as the system approach towards the respective SOC states. Our study here is for the same inequality indices, but for small time intervals along growth dynamical paths of individual researchers.

We intend to study here the inequality dynamics measured by the Gini ($g$)
and Kolkata ($k$) indices of several successful researchers (mostly
winners of international prizes, medals or awards  like Nobel, Fields,
Boltzmann, Breakthrough, highest level Stanford c-score achievers etc),
some  distinguished sociophysics researchers, along with those of a few
high level (but not so high Stanford c-score, though within ``Top 2\%")
researchers, for data up to 2022, since their recorded first publication
year. We collected their citation data of the publications  (from online
free Google Scholar, if an individual Google Scholar page exists). 
We calculate the $g$ and $k$ indices for
each year,  starting their first publication,
by taking the citation statistics today
(collected and analyzed in July-August 2023).
We extracted the values for $g$ and $k$ for
all the recorded publications of the scientist
in each overlapping five-year windows (since
the first publication), where the window
continuously shift by one year till the year
2022 (corresponding to the last central
year 2020 of the researcher) in the following
figures for each researcher. The choice of
five-year window size is found to give optimal
stability in statistics (a smaller three-year
window size did not give stability of the citation
statistics for quite a few of the scientists.)

We find, the majority of the chosen scientists
crossed the $g = k \simeq 0.82$ mark (which
we interpret here as the precursor level of the
SOC point \cite{17Manna2022}) early in their life and often
they hover just above or below but around that level
of inequality mark. Some others just touched
the precursor mark ($g = k$) once or even
multiple times and a few  remained below that
mark. For other well-known researchers
considered here, the $g = k$ mark occurs
marginally but does not cross ever. It is
to be noted that this mark of reaching
the SOC state  (beyond the $g = k \simeq$
0.82) level of inequality is for yearly
statistics (within a 5-year window which
slides yearly) and not for the overall
success measuring indices (in their
cumulative  citation statistics)
studied earlier for the citation statistics
of some distinguished researchers (see e.g.,
\cite{14Ghosh2022}), where the SOC mark is observed to be
a little higher ($g = k \simeq 0.86)$.

As mentioned earlier,  the Hirsch index ($h$) \cite{25Hirsch2005}, which gives the
highest number of publications by a  researcher,  each of
which has received equal or more than that number of citations,
does not perhaps give an excellent measure \cite{15Ghosh2023,26Yong2014} of the success
of individual researchers. It has now been clearly demonstrated \cite{15Ghosh2023}
(using the kinetic theoretical exchange
model ideas), analyzing the Scopus citation
data for the top 120,000 (within the ``Top 2\%")
Stanford cite score achievers  that statistically $h \sim \sqrt {N_c} \sim \sqrt {N_p} $,
where $N_c$  and $N_p$ denote respectively the total number of
citations and total number of publications by the researcher. This
suggests  convincingly that the  average number of citations per
paper ($N_c/N_p$), and hence $h$,  are  statistical numbers
(given by the  effective Dunbar number  \cite{28Dunbar1992,29Dunbar2010}) depending on the
community or network in which the researcher belongs \cite{15Ghosh2023,18Ghosh2021}.
We show here, extending our earlier observations (see e.g., \cite{14Ghosh2022,18Ghosh2021}),  that the
indications of success are not  reflected by the total citations
$N_c$, or for that matter by the Hirsch index $h$,  rather by the
inequalities among the citations from publication to publication.
Specifically, we show that for very successful authors, the yearly
variations (given by the statistics with overlapping 5-year windows)
in the Gini index ($g$, given by the average inequality of the
citations for the publications; $0 \le g \le 1$)  and the  Kolkata index
($k$, giving the fraction of total citations received by the top $(1-k)$
fraction of publications, $0.5 \le k \le 1$).  In particular, achieving
$g = k \simeq 0.82$ signals a precursor to the Self-Organized
Critical (SOC) state in the publication statistics. Analyzing the
citation statistics (from the open-access Google Scholar)  of
30 successful scientists throughout their recorded publication
history, starting from their first recorded publication that the very
successful among them (mostly Nobel Laureates, very high ranking Stanford c-scorers and a few others) reach and hover just
above and below
that $g = k \simeq 0.82$ mark, characteristic of the SOC state
($k = 0.82$ means 82\% citations come from 18\% publications).
Others remain below that (SOC) level of
extreme inequality in publication statistics.

\section{Socio-statistical Inequality and Its Measures}
\begin{figure}[H]
    \centering
    \includegraphics[width=0.7\textwidth]{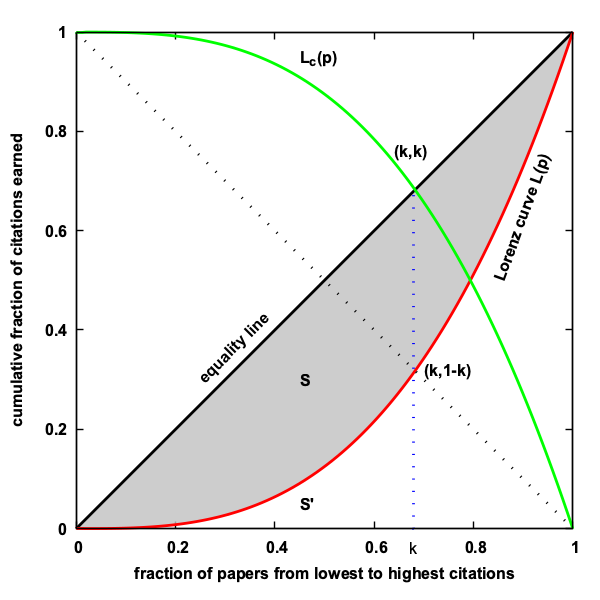}
    \caption{The Lorenz curve, represented by $L(p)$ in red, denotes the
cumulative proportion of total citations possessed by a  fraction
($p$) of papers, when organized in ascending order of citation counts.
Conversely, the black dotted line indicates perfect equality, where
each paper receives an equal number of citations. The Gini index ($g$)
is computed from the area ($S$) between the Lorenz curve and the
equality line (the shaded region), normalized by the total area under
the equality line ($S + S' = 1/2$). The Kolkata index ($k$) is
obtained  by locating the fixed point of the complementary Lorenz
function ($L_c$), defined as $L_c(p) \equiv 1 - L(p)$:  $L_c(k) = k$.
By geometry, the value of
$k$ gives  the proportion of total citations
owned or possessed by $(1 - k)$ fraction of
the top cited papers.}
    \label{fig1-lorentz.png}
\end{figure}

In 1905  American economist Lorenz \cite{2Chakrabarti2006,3Castelano2009} developed the Lorenz
curve, a graphical representation of the distribution of wealth in a
society. To construct this curve (illustrated by the red curve in
Figure \ref{fig1-lorentz.png}), one organizes the society's population in ascending order
of their wealth and then plots the cumulative fraction of wealth,
denoted as $L(p)$, held by the poorest $p$ fraction of individuals.
One can similarly plot the cumulative fraction of citations against the
fraction of papers that attracted those many citations. As indicated
in Fig. \ref{fig1-lorentz.png}, 
the Gini index is calculated from the area between the equality
line and the Lorenz curve, divided by the area (1/2) below the
equality line for normalization. As such, $g = 0$ signifies perfect
equality and $g = 1$ corresponds to extreme inequality.
The Kolkata index
$k$ is given by the fixed point of the Complementary  Lorentz function
$L_c (p) \equiv  1 - L(p)$. As such, $k$ gives the fraction of
citations attracted by the top cited $k$ fraction of papers and $k
=0.5$ means perfect equality, while extreme inequality corresponds to
$k = 1$.

\subsection{Landau-like expansion of $L(p)$ and $g, k$ approximate relation}
A minimal expansion \cite{29Joseph2022} of the Lorenz
function $L(p)$, employing a Landau-like
expansion of free energy, suggests
$L(p) = Ap + Bp^2, A > 0, B >0, A + B = 1$.
This gives $L (0) = 0$ and $L (1) = 1$
(with $B$ = 0, the Lorenz function can
represent only the equality line in Figure \ref{fig1-lorentz.png}).

One can then calculate $g = 1-2\int_0^1 L(p)dp$,
giving $A = 1 -3g$ and $B = 3g$.  Since $L(k) = 1- k = Ak + Bk^2$, one can obtain a quadratic
equation involving $g$ and $k$.  An approximate
solution of it, in the $g \rightarrow 0$ limit gives
\begin{equation}
    k = 1/2 + C*g,
    \label{eqn1}
\end{equation}
where $C=3/8$ \cite{29Joseph2022} suggesting that
$g =k$ will occur at the Pareto value  $k = 0.80$.
We will see here a little deviation in the value of
the  constant $C$ in the relation (\ref{eqn1}), for all the
reported observations.

\section{Inequality data analysis from Google Scholar}
We collect the citation data for all the
recorded publications in each year since
the first entry in the record for thirty
successful researchers having individual
Google Scholar page and having minimum
and maximum number of total publications
$N_p$ = 127 and 2954,  minimum and maximum
number of total citations $N_c$ = 5769 and
463382, minimum and maximum values of
Hirsch index $h$ = 22 and 328, respectively
for all those selected researchers. 
We considered three Nobel prize winners in each of the science subjects:
Physics (H. Amano, B. Josephson, A. B. McDonald),  Chemistry (R.
Henderson, J. Frank, J.-P. Sauvage),  Physiology \& Medicine (M. Houghton,
G. L. Semanza, S. Yamanka), and Economics (A. Banerjee, W. Nordhaus, J.
Stiglitz). Two Fields medalists (Mathematics; S. Smalle, E. Witten), two
Boltzmann award winners (Statistical Physics; D. Dhar, H. E. Stanley), two
Breakthrough Prize winners (Physics; C. Kane, A. Sen), three of the
top-most cite-scorers in the Stanford Scopus c-score list (M. Graetzel, R.
C. Kessler and Z. L. Wang; considered for $h$-index statistics in \cite{15Ghosh2023}), and six
well-known contributors in Econophysics and Sociophysics:  W. Brian Arthur
(known for ``El Farol Bar Problem" of  
minority choice, see e.g., \cite{30Challet2005}), B. K.
Chakrabarti (one of the ``Fathers of
Econophysics” \cite{31Jovanovic2017,32Schinckus2018}),  R. I. M. Dunbar (known
for Dunbar’s number of social connectivity,
see e.g., \cite{33wiki}), S. Galam (considered  Pioneer
of Sociophysics, see e.g., contributions in
this Special Issue \cite{34Galam70}), R. Mantegna (one of
the ``Fathers of Econophysics” \cite{31Jovanovic2017,32Schinckus2018}), V. M.
Yakovenko (pioneer of kinetic exchange models
of income/wealth distributions, see e.g., \cite{35Yakovenko2009}).
We considered three of the highest-ranked
Stanford Cite-Scorers for 2022 (M. Graetzel,
R. C. Kessler and Z. L. Wang  \cite{36Ioannidis2022}), and for
comparison, we also considered three  lower rank holders
 of the same ``Top 2\% Stanford Cite-Scores" (I. Fofana, U. Sennur and N.
Tomoyuki \cite{36Ioannidis2022}).

For studying the growth of inequality in the citation-statistics of each
of these researchers, we select a 5-year window, starting earliest
publication, and note the present-day citations of each of these
publications. We then construct the Lorenz function (see Fig. \ref{fig1-lorentz.png}) and
extract the $g$ and $k$  indices as described the last section. We
associate the $g$ and $k$ values with the middle year of the respective
5-year window and by one and shift the window by one year and get the
values of the inequality indices for each of the successive years
up to 2020 (considering data up to 2022). These are shown in the following Figs. 2-6.


We can see from the Figs. 2-6, for all the
above-mentioned 30 scientists that for many of
them (mostly Nobel Prize winners and highest
rank c-scorers), the Gini index $g$ value
goes over the Kolkata index $k$ value in one
(or multiple years) by crossing the
$k = g \simeq 0.82$ line (see the
corresponding insets). These crossings of the
indices (at values above 0.80 value) clearly
indicates large inequalities and entering in
to the Self-Organized Critical (SOC) state
of the citation statistics of these
scientists \cite{17Manna2022}.

\begin{figure}[H]
    \centering
    \includegraphics[width=0.95\textwidth]{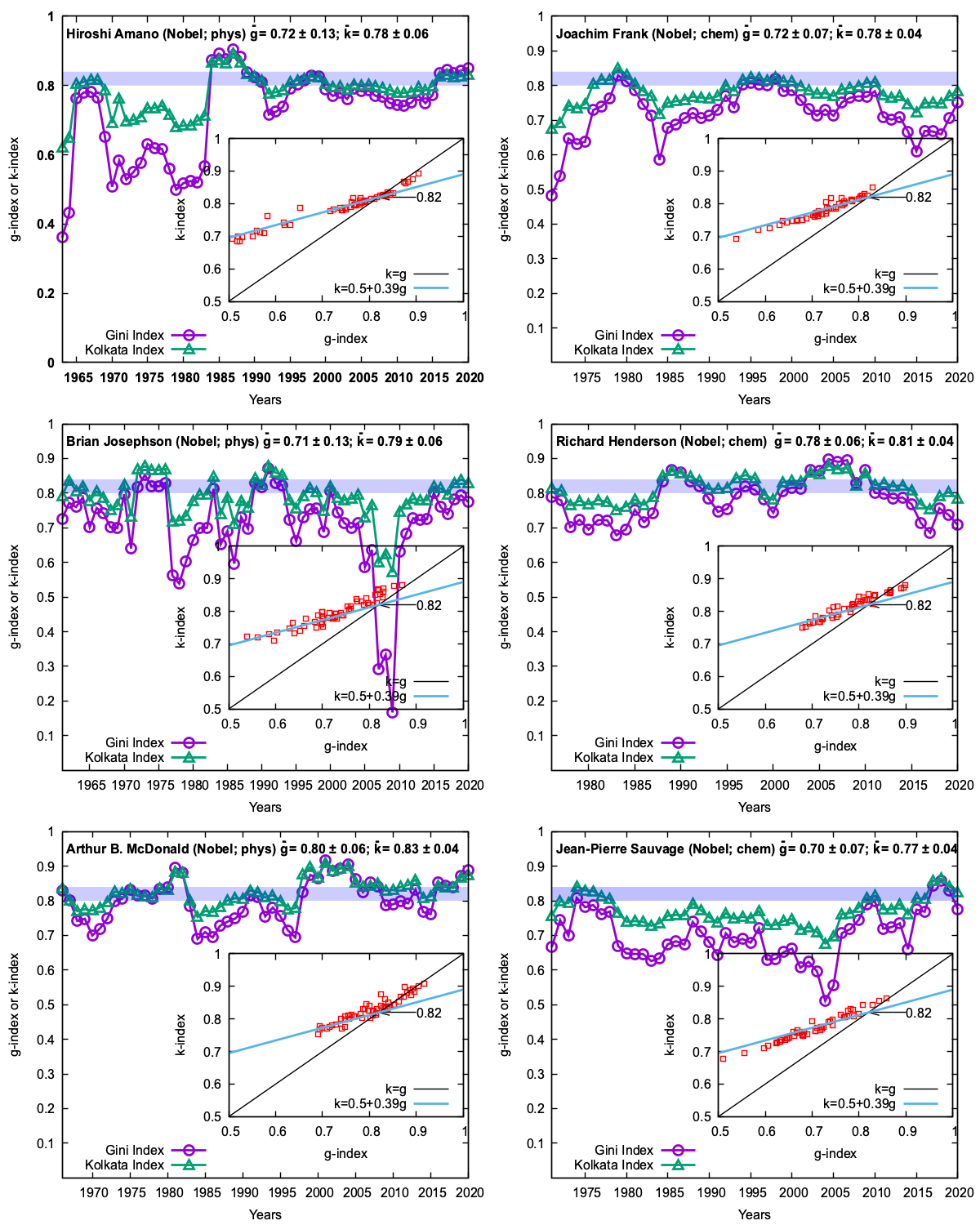}
    \caption{Yearly variations of the citation inequality indices,  Gini ($g$) and
Kolkata ($k$), for 3 Nobel prize winners in Physics and 3 in Chemistry.
The indices are calculated using the present citation data for the
publications within a 5-year window, starting from first recorded one in
Google Scholar, and the window sliding by one year. The corresponding
year shown is mid year of the window until 2022 (shown for year 2020 for
the last 5-year window). The $g$ value crossing above (and coming down)
the $k$ value marks
the precursor of onset (leaving) the SOC state with time. The inset shows
the plot of $k$ vs. $g$ over the entire career of the scientist. It
fits well with the linear (Landau-like) relationship $k = 1/2 + 0.39g$,
suggesting a crossing SOC precursor point at $k = g = 0.82 \pm 0.02$.}
\label{phys-chem.png}
\end{figure}

\begin{figure}[H]
    \centering
    \includegraphics[width=0.95\textwidth]{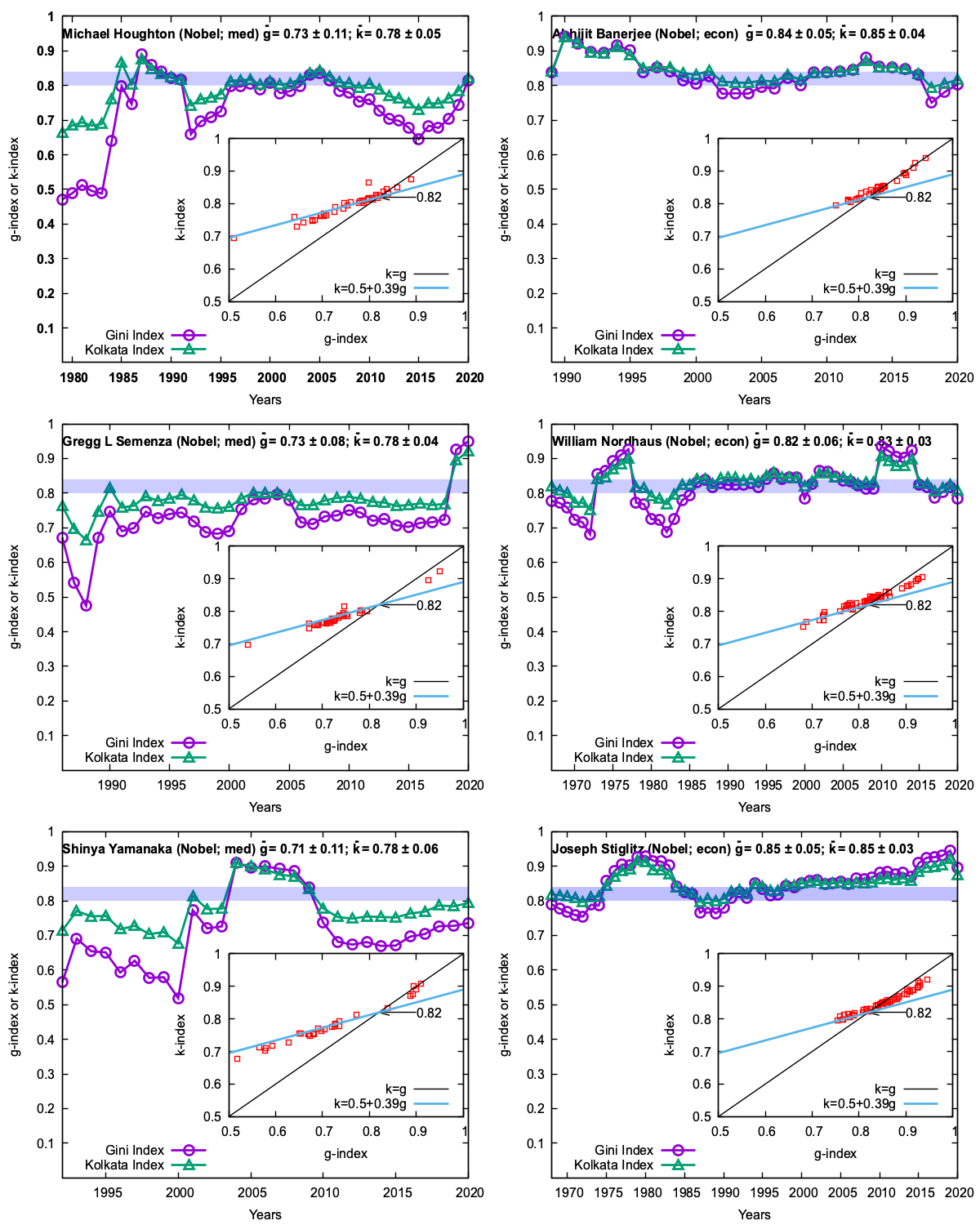}
    \caption{Yearly variations of the citation inequality indices,  Gini ($g$) and
Kolkata ($k$), for 3 Nobel prize winners in Physiology-Medicine  and 3 in
Economics. The indices are calculated using the present citation data for
the publications within a 5-year window, starting from the first recorded one
in Google Scholar, and the window sliding by one year. The corresponding
year shown is mid year of the window until 2022 (shown for the year 2020 for
the last 5-year window). The $g$ value crossing above (and coming down)
the $k$ value marks the precursor of onset (leaving) the SOC state with
time. The inset shows the plot of $k$ vs. $g$ over the entire career
of the scientist. It fits well with the linear (Landau-like) relationship
$k = 1/2 + 0.39g$, suggesting a crossing SOC precursor point at $k = g =
0.82 \pm 0.02$.}
    \label{bio-econ.png}
\end{figure}

\begin{figure}[H]
    \centering
    \includegraphics[width=0.95\textwidth]{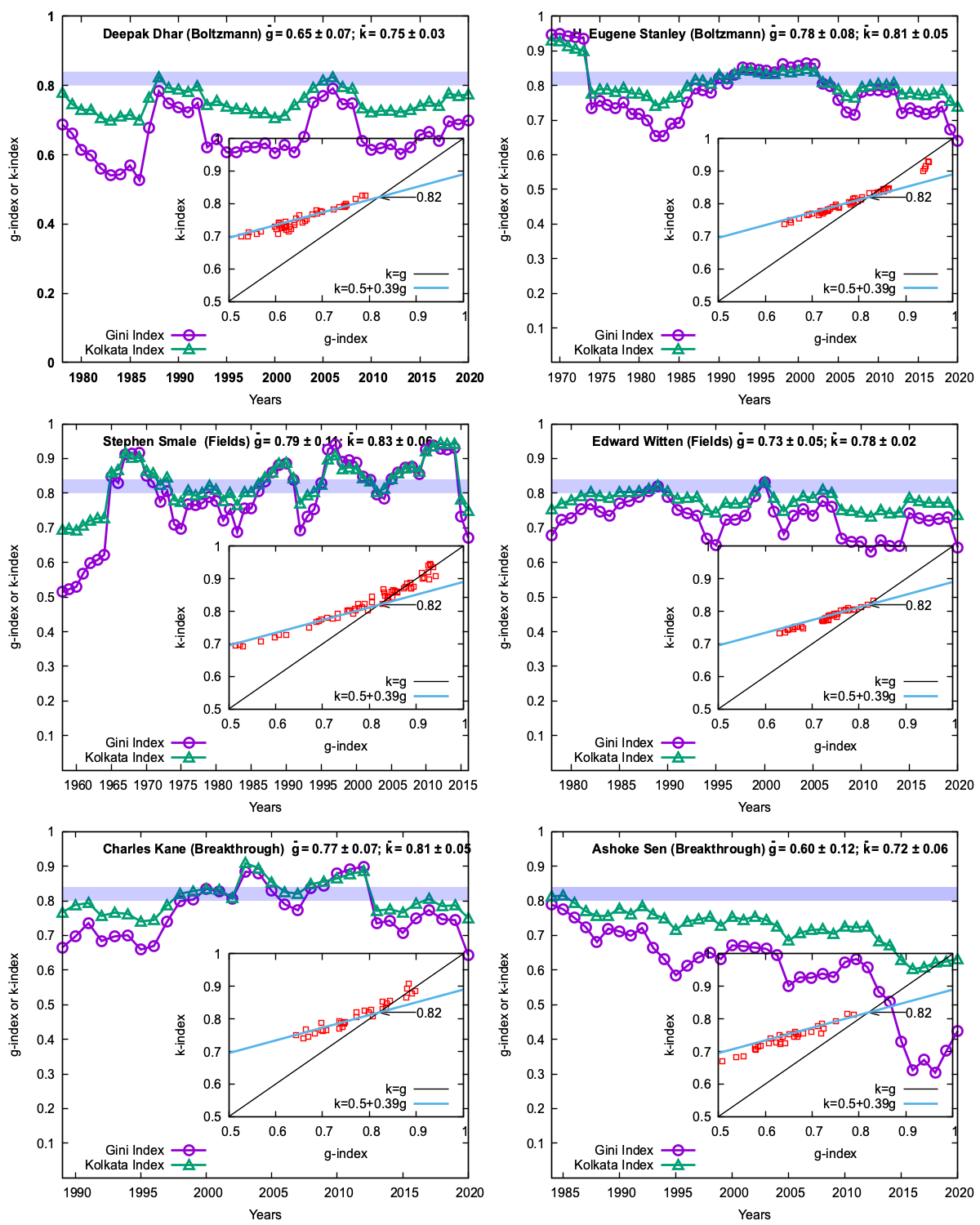}
    \caption{Yearly variations of the citation inequality indices,  Gini ($g$) and
Kolkata ($k$), for 2  winners each of Fields Medal (Mathematics),
Boltzmann Prize (Statistical Physics) and Breakthrough Prize (Physics).
The indices are calculated using the present citation data for the
publications within a 5-year window, starting from first recorded one in
Google Scholar, and the window sliding by one year. The corresponding
year shown is mid year of the window until 2022 (shown for year 2020 for
the last 5-year window). The $g$ value crossing above (and coming down)
the $k$ value marks the precursor of onset (leaving) the SOC state with
time. The inset shows the plot of $k$ vs. $g$ over the entire career
of the scientist. It fits well with the linear (Landau-like) relationship
$k = 1/2 + 0.39g$, suggesting a crossing SOC precursor point at $k = g =
0.82 \pm 0.02$.}
    \label{prize-winner.png}
\end{figure}

\begin{figure}[H]
    \centering
    \includegraphics[width=0.95\textwidth]{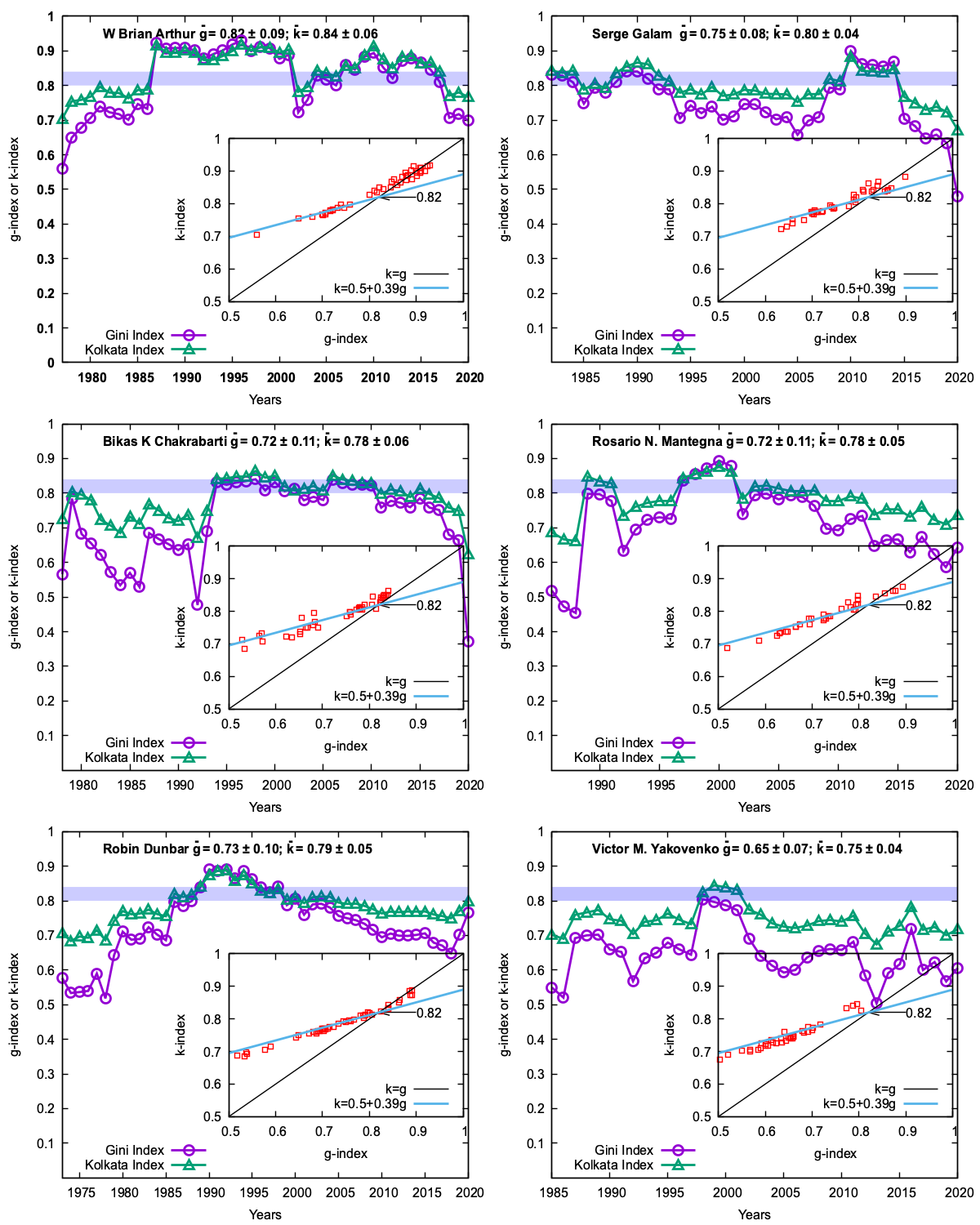}
    \caption{Yearly variations of the citation inequality indices,  Gini ($g$) and
Kolkata ($k$), for 6 distinguished researchers in Econophysics and
Sociophysics. The indices are calculated using the present citation data
for the publications within a 5-year window, starting from first recorded
one in Google Scholar, and the window sliding by one year. The
corresponding year shown is mid year of the window until 2022 (shown for
year 2020 for the last 5-year window). The $g$ value crossing above (and
coming down) the $k$ value marks the precursor of onset (leaving) the SOC
state with time. The inset shows the plot of $k$ vs. $g$ over the
entire career of the scientist. It fits well with the linear (Landau-like)
relationship $k = 1/2 + 0.39g$, suggesting a crossing SOC precursor point
at $k = g = 0.82 \pm 0.02$.}
    \label{no-prize.png}
\end{figure}

\begin{figure}[H]
    \centering
    \includegraphics[width=0.95\textwidth]{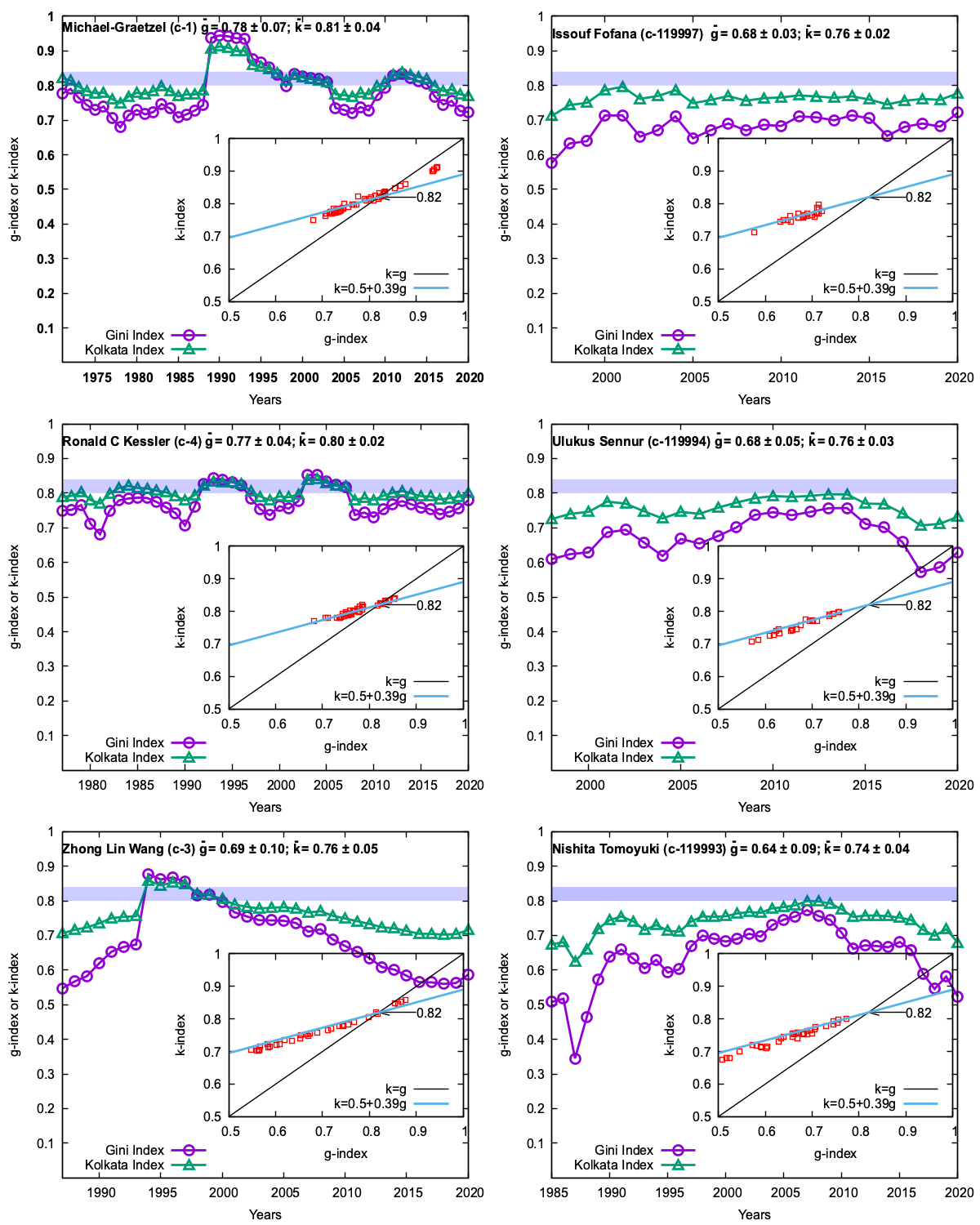}
    \caption{Yearly variations of the citation inequality indices,  Gini ($g$) and
Kolkata ($k$), for 3 top-most Stanford Cite-Scores and 3 lower rank entries from
the ``Top 2\% Stanford Cite-Scores" \cite{15Ghosh2023,36Ioannidis2022}. The indices are calculated using the
present citation data for the publications within a 5-year window,
starting from first recorded one in Google Scholar, and the window sliding
by one year. The corresponding year shown is mid year of the window
until 2022 (shown for year 2020 for the last 5-year window). The $g$ value
crossing above (and coming down) the $k$ value marks the precursor of
onset (leaving) the SOC state with time. The inset shows the plot of $k$
vs. $g$ over the entire career of the scientist. It fits well with the
linear (Landau-like) relationship $k = 1/2 + 0.39g$, suggesting a crossing
SOC precursor point at $k = g = 0.82 \pm 0.02$.}
    \label{c-score.png}
\end{figure}

\begin{table}[H]
\centering
\caption{Consolidated inequality index ($g, k$)
results for the citation statistics (from Figs. 2-6) of the 30 chosen science researchers
(including 12 Nobel prize winners, 2 Fields
Medalists, 2 Boltzmann Award winners, 2
Breakthrough Prize winners, 6 distinguished
Sociophysics and Econophysics researchers,
3 from the top and 3 from the bottom of the
``Top 2\% Stanford Cite-Score Scientists" (2022 list). NP(P) means Nobel Prize in
Physics,  NP(C) means Nobel Prize in Chemistry,
NP(M) means Nobel Prize in Physiology or
Medicine and NP(E) means Nobel Prize in
Economics, FM means Fields Medal in
Mathematics,  BA means Boltzmann Award in
Statistical Physics, BP(P) means
Breakthrough Prize in Physics,  FEP means ``Father of Econophysics" \cite{31Jovanovic2017,32Schinckus2018}, EFBP means ``El Farol Bar problem" (see e.g., \cite{30Challet2005}), 
DN means ``Dunbar Number" (see e.g., \cite{33wiki}),
FSP means ``Father of Sociophysics" (see e.g.,
contributions in this  Spl. issue \cite{34Galam70}),
PKEM means Pioneer in Kinetic Exchange
Modeling of Wealth Distribution \cite{35Yakovenko2009},
SCS-x means Stanford Cite Score rank
(x denoting the rank) among the ``Top 2\%"
scientists in 2022 \cite{36Ioannidis2022}.
\newline
}

\resizebox{\textwidth}{!}{\begin{tabular}{|l|l|c|c|c|c|c|c|c|c|}
\hline
  & &   &  &     \multicolumn{5}{c|}{Inequality Indices: Hirsch ($h$), Gini ($g$), Kolkata ($k$)}         &  $g=k$ line \\
  \cline{5-9}
Researcher  &Award/Prize& $N_P$  & $N_C$  & $h$  & $g$  & $k$  &  $g$  &  $k$ & crossed \\
Name &/Known for  & &  &   & (overall) & (overall) & (Yearly Av.) & (Yearly-Av.) &  near $0.82?$ \\
\hline
H Amano &  NP(P)        & 2161 & 57281  & 106 & 0.84 & 0.83 & 0.72$\pm$0.13 & 0.78$\pm$0.06 & Yes \\
\hline
B Josephson  &   NP(P)         & 127  & 11685  & 22  & 0.94 & 0.92 & 0.71$\pm$0.13 & 0.79$\pm$0.06 & No \\
\hline
AB McDonald   &    NP(P)         & 437  & 25111  & 53 & 0.91 & 0.88 & 0.80$\pm$0.06 & 0.83$\pm$0.04 & Yes \\
\hline
J Frank  &  NP(C)   & 686  & 50518  & 116  & 0.77 & 0.80 & 0.72$\pm$0.07 & 0.78$\pm$0.04 & No \\
\hline
R Henderson  &  NP(C)  & 267  & 31822  & 65  & 0.85 & 0.84 & 0.78$\pm$0.06 & 0.81$\pm$0.04  & Yes \\
\hline
JP Sauvage  &  NP(C) & 655  & 61572  & 114 & 0.70 & 0.76 & 0.70$\pm$0.07 & 0.77$\pm$0.04  & Marginally \\
\hline
M Houghton  &  NP(M)  & 529  & 59029  & 102 & 0.85 & 0.84 & 0.73$\pm$0.11 & 0.78$\pm$0.05 & Yes\\
\hline
GL Semenza  &  NP(M)   & 682  & 192246 & 196 & 0.80 & 0.81 & 0.73$\pm$0.08 & 0.78$\pm$0.04 & Yes \\
\hline
S Yamanaka  &  NP(M)    & 345  & 124106 & 125 & 0.85 & 0.84 & 0.71$\pm$0.11 & 0.78$\pm$0.06 & Yes\\
\hline
A Banerjee  &   NP(E)    & 524  & 79076  & 106 & 0.86 & 0.86 & 0.84$\pm$0.05 & 0.85$\pm$0.04 & Yes \\
\hline
W Nordhaus  &  NP(E)    & 647  & 101219 & 124 & 0.87 & 0.86 & 0.82$\pm$0.06 & 0.83$\pm$0.03 & Yes \\
\hline
J Stiglitz    &     NP(E)       & 2408 & 364237 & 235 & 0.89 & 0.87 & 0.85$\pm$0.05 & 0.85$\pm$0.03 & Yes\\
\hline
D Dhar  &      BA           & 209  & 8299   & 44  & 0.76 & 0.80 & 0.65$\pm$0.07 & 0.75$\pm$0.03 & No\\
\hline
HE Stanley   &      BA, FEP      & 2070 & 225169 & 204 & 0.83 & 0.83 & 0.78$\pm$0.08 & 0.81$\pm$0.05 & Yes\\
\hline
S Smale   &    FM    & 346  & 48084  & 85 & 0.87 & 0.85 & 0.79$\pm$0.11 & 0.83$\pm$0.06  & Yes\\
\hline
E Witten  &  FM, BP(P)       & 620  & 242911 & 206 & 0.79 & 0.81 & 0.73$\pm$0.05 & 0.78$\pm$0.02 & Marginally\\
\hline
C Kane  &   BP(P)     & 189  & 80714  & 75 & 0.88 & 0.87 & 0.77$\pm$0.07 & 0.81$\pm$0.05 & Yes\\
\hline
A   Sen   &    BP(P)       & 401  & 37065  & 103 & 0.69 & 0.76 & 0.60$\pm$0.12 & 0.72$\pm$0.06 & No \\
\hline
WB Arthur  &  EFBP      & 196  & 52545  & 56  & 0.91 & 0.89 & 0.82$\pm$0.09 & 0.84$\pm$0.06 & Yes \\
\hline
BK Chakrabarti  &     FEP            & 390  & 12596  & 47 & 0.81 & 0.82 & 0.72$\pm$0.11 & 0.78$\pm$0.06 & Marginally \\
\hline
RIM Dunbar &     DN             & 857  & 85486  & 141 & 0.79 & 0.81 & 0.73$\pm$0.10 & 0.79$\pm$0.05 & Yes \\
\hline
S Galam   &       FSP         & 252  & 8828   & 42 & 0.81 & 0.83 & 0.75$\pm$0.08 & 0.80$\pm$0.04  &  Yes \\
\hline
RN Mantegna  &    FEP         & 259  & 28561  & 68  & 0.84 & 0.84 & 0.72$\pm$0.11 & 0.78$\pm$0.05  &  Yes \\
\hline
VM Yakovenko  & PKEM  & 171  & 9076   & 44 & 0.73 & 0.78 & 0.65$\pm$0.07 & 0.75$\pm$0.04 & No \\
\hline
M Graetzel  &  SCS-1  & 2282 & 463382 & 295 & 0.82 & 0.82  & 0.78$\pm$0.07 & 0.81$\pm$0.04 & Yes\\
\hline
RC Kessler  &  SCS-4  & 1829 & 523835 & 328 & 0.83 & 0.83 & 0.77$\pm$0.04 & 0.80$\pm$0.02 & Yes\\
\hline
ZL Wang  &  SCS-3   & 2954 & 394080 & 299 & 0.71 & 0.77 & 0.69$\pm$0.10 & 0.76$\pm$0.05 &  Yes\\
\hline
I Fofana  & SCS-119997  & 353  & 5759   & 39  & 0.73 & 0.78 & 0.68$\pm$0.03 & 0.76$\pm$0.02 & No\\
\hline
N Tomoyuki &  SCS-119993    & 304  & 9073   & 51 & 0.75 & 0.79 & 0.64$\pm$0.09 & 0.74$\pm$0.04 & No \\
\hline
U Sennur  & SCS-119994  & 455  & 18987  & 63  & 0.73 & 0.78  & 0.68$\pm$0.05 & 0.76$\pm$0.03 & No\\
\hline
\end{tabular}}
\label{table1}
\end{table}

Although the study of the time variations
of the Gini ($g$) and Kolkata ($k$) indices
(as shown in Figs. 2-6) and checking if $g$
value ever goes over the $k$ value by
crossing the $k$ vs. $g$ line (as shown in
the respective insets) is indispensable
for detecting if the SOC state has arrived
or not, one can also have an easy (but only
approximate) indication of the SOC state
by looking at the ratio $R$ of the citation
number $n_C^{max}$ of the highest cited
paper and the effective Dunbar number $D$
given by the average citation $N_C/N_p$ of
the researcher. In Table \ref{table2}, we precisely
compare these $R = n_C^{max}/D$ values
(where $D = N_C/N_P$) and see how its
higher values compare with the observation
of SOC (when $k$ vs. $g$ line is crossed
affirmatively). We find, for $R \ge 40$
more than $94\%$ cases correspond to SOC level.

\begin{table}[H]
\centering
\caption{A rough indicator $R = n_C^{max}/D$, where
the effective Dunbar number $D = N_C/N_P$
($N_C$ denotes the total number of citations for
$N_P$ papers by the researcher) and
$n_C^{max}$ denotes the citation of the
most-cited paper by the researcher, to
check if the researcher has achieved the
SOC level or not. We find, for $R \geq 40$ the corresponding researchers clearly belong to the SOC level ($94\%$ success rate).\newline
}

\resizebox{\textwidth}{!}{\begin{tabular}{|l|l|l|l|l|l|l|l|l|l|}
\hline
Researcher & Award/ & $N_P$ & $N_C$ & $n_C^{max}$ & $h$ & $D=$ &  $R=$ & SOC level & Comments\\
Name       & Known &    &   &          & & $N_C/N_P$ &  $n_C^{max}/D$ & achieved  &        \\
          & for  &   &   &          &  &  &  &  (Table I) &       \\
\hline
H Amano        & NP(P)      & 2161 & 57281  & 3154  & 106 & 27  & 119 & Yes        & \multirow{31}{*}{} \\ \cline{1-9}
B Josephson    & NP(P)      & 127  & 11685  & 6554  & 22  & 92  & 71  & No         &  Out of the                 \\ \cline{1-9}
AB McDonald    & NP(P)      & 437  & 25111  & 5375  & 53  & 57  & 94  & Yes        &   eighteen               \\ \cline{1-9}
J Frank        & NP(C)      & 686  & 50518  & 2299  & 116 & 74  & 31  & No         &   researchers                \\ \cline{1-9}
R Henderson    & NP(C)      & 267  & 31822  & 3681  & 65  & 119 & 31  & Yes        &  with $R \ge 40$,                  \\ \cline{1-9}
JP Sauvage     & NP(C)      & 655  & 61572  & 1805  & 114 & 94  & 19  & Marginally &   one failed                 \\ \cline{1-9}
M Houghton     & NP(M)      & 529  & 59029  & 9952  & 102 & 112 & 89  & Yes        &   in achieving                 \\ \cline{1-9}
GL Semenza     & NP(M)      & 682  & 192246 & 12229 & 196 & 282 & 43  & Yes        &    the SOC level                \\ \cline{1-9}
S Yamanaka     & NP(M)      & 345  & 124106 & 30735 & 125 & 360 & 85  & Yes        &   (crossing the                 \\ \cline{1-9}
A Banerjee     & NP(E)      & 524  & 79076  & 9254  & 106 & 151 & 61  & Yes        &  $k = g \simeq 0.82$             \\ \cline{1-9}
W Nordhaus     & NP(E)      & 647  & 101219 & 19605 & 124 & 156 & 125 & Yes        &   line in                 \\ \cline{1-9}
J Stiglitz     & NP(E)      & 2408 & 364237 & 23844 & 235 & 151 & 158 & Yes        &   Figs. 2-6;                  \\ \cline{1-9}
D Dhar         & BA         & 209  & 8299   & 1182  & 44  & 40  & 30  & No         &   see Table I).                 \\ \cline{1-9}
HE Stanley     & BA, FEP    & 2070 & 225169 & 14348 & 204 & 109 & 132 & Yes        &   $R \ge 40$                 \\ \cline{1-9}
S Smale        & FM         & 346  & 48084  & 7912  & 85  & 139 & 57  & Yes        &  therefore                  \\ \cline{1-9}
E Witten       & FM, BP(P)  & 620  & 242911 & 14380 & 206 & 392 & 37  & Marginally &  indicates                  \\ \cline{1-9}
C Kane         & BP(P)      & 189  & 80714  & 19504 & 75  & 427 & 46  & Yes        &   SOC level                  \\ \cline{1-9}
A   Sen        & BP(P)      & 401  & 37065  & 1443  & 103 & 92  & 16  & No         &   for the                 \\ \cline{1-9}
WB Arthur      & EFBP       & 196  & 52545  & 15227 & 56  & 268 & 57  & Yes        &   researcher                 \\ \cline{1-9}
BK Chakrabarti & FEP        & 390  & 12596  & 730   & 47  & 32  & 23  & Marginally &  with more                  \\ \cline{1-9}
RIM Dunbar     & DN         & 857  & 85486  & 5312  & 141 & 100 & 53  & Yes        &   than 94\%                 \\ \cline{1-9}
S Galam        & FSP        & 252  & 8828   & 653   & 42  & 35  & 19  & Yes        &    success rate.                \\ \cline{1-9}
RN Mantegna    & FEP        & 259  & 28561  & 5796  & 68  & 110 & 53  & Yes        &                    \\ \cline{1-9}
VM Yakovenko   & PKEM       & 171  & 9076   & 920   & 44  & 53  & 17  & No         &                    \\ \cline{1-9}
M Graetzel     & SCS-1      & 2282 & 463382 & 35789 & 295 & 203 & 176 & Yes        &                    \\ \cline{1-9}
RC Kessler     & SCS-4      & 1829 & 523835 & 35079 & 328 & 286 & 122 & Yes        &                    \\ \cline{1-9}
ZL Wang        & SCS-3      & 2954 & 394080 & 8120  & 299 & 133 & 61  & Yes        &                    \\ \cline{1-9}
I Fofana       & SCS-119997 & 353  & 5759   & 333   & 39  & 16  & 20  & No         &                    \\ \cline{1-9}
N Tomoyuki     & SCS-119993 & 304  & 9073   & 467   & 51  & 30  & 16  & No         &                    \\ \cline{1-9}
U Sennur       & SCS-119994 & 455  & 18987  & 1198  & 63  & 42  & 29  & No         &                    \\ \cline{1-9}
\hline
\end{tabular}}
\label{table2}
\end{table}

\section{Summary and Discussions}
Our earlier analysis \cite{15Ghosh2023} of the Scopus citation  data for  the
120000 top Stanford Cite-Score scientists showed that the Hirsch 
index $h \sim {\sqrt N_c} \sim {\sqrt N_p}$ , where $N_c $ and $N_p $
denote respectively the total number of citations and the total
number of publications by the researcher. This, in turn, says that
the average number of citations per paper ( $N_c/N_ p)$, and hence
$h$, are statistical numbers (determined by the effective Dunbar
number \cite{28Dunbar1992,33wiki}) of the community or network (coauthors and followers)
in which the researcher belongs \cite{15Ghosh2023,18Ghosh2021}. Indeed the anticipated  increase
of  research impact  through collaboration (by increasing the number of
coauthors) have been studied in \cite{37Katz1997}, by looking at the average value
of the community Dunbar number or $N_c/N_p$. Also, detailed study
from Google Scholar data on the relation between Hirsch index  of
individual scientists with their  average number of co-authors per paper
has been reported in ref. \cite{38Arnaboldi2016}.  Our study here shows that Hirsch index
can not  be a good measure of success for the researchers (even in
Table \ref{table1}; the highest $h = 328$ does not correspond to a Nobel Prize
winner, while the least one with $h = 22$ do).

In an earlier work \cite{18Ghosh2021}, we proposed
that the citation inequality indices
Gini $(0 \le g \le 1)$ and Kolkata
$(0.5 \le k \le 1)$ might give better
measures of success of the scientist
(not $N_c$ or $h$) and perhaps $g$ and
$k$ both approach to equality at
$g = k \simeq 0.86$ for successful
researchers. It may
be mentioned here that we used there the
entire citation data (over all the years)
to get the Lorenz curve and the overall
values of $g$ and $k$ of the researcher,
and this gave a little higher value of
$g = k \simeq 0.86$ point. Indeed, our
numerical study \cite{17Manna2022} of the overall or
cumulative  inequality statistics of the
avalanches or cluster sizes in some
well-studied and well-established
Self-Organized Critical (SOC) models
also suggested the arrival of the
equality point of the avalanche size
inequality indices ($g = k \simeq 0.86$)
just appears as a precursor of the SOC point
of the respective sand-pile or SOC models.
In other words, as mentioned already, the SOC
points in sand pile models (like BTW, Manna,
etc) of physics signifies a critical state
where sand grain avalanches of all sizes occur
following a power law distribution. As shown in
Ref.\cite{17Manna2022}, even in these physics SOC models, the
inequality statistics (indices Gini \& Kolkata)
corresponding to the avalanche size statistics
reach similar values for the inequality indices
of the unequal citations (considered here
equivalent to the sand mass avalanches in sand
piles).

We analyzed here the citation data
for all the recorded publications in each
year since the first entry in the record
for the chosen  30 successful scientists,
each having an individual Google Scholar page.
They have the minimum and maximum number of
total publications $N_p$ = 127 and 2954,
and minimum and maximum number of total
citations $N_c$ = 5769 and 463382,
respectively. For studying the growth of
inequality in the citation statistics if
each of these scientists, we select 5-year
windows, where the central year of each
window moves every year. We constructed
the Lorenz functions for each of these
windows (see Fig. \ref{fig1-lorentz.png}) and extract the
yearly values (corresponding to the central
year of the window) of $g$ and $k$ indices.
We have plotted these yearly $g$ and $k$
values for all the working years, starting
the recorded first year and for the third year
from there and continued for successive years
up to 2022 (by considering data up to 2022)
for each of these chosen 30 scientists.
These are then shown in Figs. 2-6. The
insets in each Fig. show the corresponding
plot of $k$ vs. $g$ (disregarding the yearly
sequence). These plots in all 30 cases of the
researchers show very good linear fit to
$k = 0.5 + 0.39g$ (cf. eqn. (\ref{eqn1})), as obtained
approximately using a (Landau-like) minimal
polynomial expansion  of the Lorenz function
(see section II.A). The insets also show the
actual or extrapolated (precursor of sand-pile
SOC) point at $k = g = 0.82 \pm 0.02$. As we
can see from Figs. 2-6, for 10 of the 12 Nobel
Prize winners, several of the other
International prize winners are considered here,
well known  Sociophysicists, Econophysicists,
and all 3 of the highest rank Stanford Cite-
Scorers, the crossing(s) of $k$ vs. $g$ (often
at multiple years), do take place convincingly.
The same is also true (often marginally), for
several others. The 3 lower rank (yet from
the ``Top 2\%") Stanford Cite-Scorers 
did not come up to $g = k$ point.  There are
of course a few notable anomalies  in this
analysis of the data set; e.g., B. Josephson,
J. Frank (both Nobel Laureates), D. Dhar
(Boltzmann Award winner) and A. Sen
(Breakthrough Prize winner) do not fit this
picture of clearly reaching the SOC point.
These anomalies indicate perhaps  some
shortcomings of this kind of analysis. On
the other hand, noting that out of 27 of the
researchers have chosen here (neglecting the 3
lower rank, though from the ``Top 2\%", Stanford
Cite-Scorers), the clear evidence of SOC are
seen for 19 (neglecting the ``no" and ``marginal"
entries in the last column of Table \ref{table1} for these
27 researchers), indicating a success
rate more than 70\% for identifying the
outstanding researchers. In Table \ref{table2},
we give a simple (though rough) indicator
$R = n_C^{max}/D$ (where $n_C^{max}$ denotes
the maximum citation of any paper and
$D$ the effective Dunbar number of the
researcher) to check if the researcher
has achieved the SOC level or not. We see that the SOC level is achieved for $R \geq 40$, with more than 94\% coincidence rate.

In summary, as statistically the Hirsch index $h$ of a prolific researcher
grows
with the total citations $N_c$ as
$h = 0.5 \sqrt{N_c}$ \cite{15Ghosh2023} and $N_c$
grows linearly with the total number
$N_p$ of publications by the researcher,
$N_c = D N_p$ (see \cite{15Ghosh2023,18Ghosh2021}), where
the effective Dunbar number
$D$ ($\sim 75$ \cite{15Ghosh2023}) of the network
community in which the scientist belongs,
$h$ and $N_c$ can only give some average
measures of success. In fact, very well
appreciated members of the community can
in principle  have uniformly high citations
of order $D$ for each of their publications
and hence $h \ge D \simeq$ 75. Though such
uniformly appreciated or cited scientists
will have very low values of Gini
($g \simeq$ 0) and Kolkata ($k \simeq 0.5$)
index values. Our study here shows, notwithstanding some
anomalies, most successful researchers have
large fluctuations in the citations of one or more of their publications
(presumably due to uneven but accurate
appreciations from the usual Dunbar network or
community and also perhaps from outside the usual
Dunbar community), which do not increase  directly the $D$ or
$h$ values, but lead to larger values of
their inequality indices $g$ and $k$, which
may then hover around the SOC level value
$g = k \simeq 0.82$, a little above the
Pareto  value ($k$ = 0.80).

\section*{Acknowledgement}
 We are thankful to  Soumyajyoti Biswas, Arnab Chatterjee, Arnab Das
for careful readings of the manuscript
and for useful criticisms and suggestions. BKC is grateful to the Indian National Science Academy for their Senior Scientist Research Grant.



\end{document}